# Asking Biological Questions of Physical Systems: the Device Approach to Emergent Properties

**Bob Eisenberg**

Department of Applied Mathematics
Illinois Institute of Technology
USA

Department of Physiology and Biophysics
Rush University
USA

bob.eisenberg@gmail.com

January 17, 2018

## Abstract

Life occurs in concentrated 'Ringer Solutions' derived from seawater that Lesser Blum studied for most of his life. As we worked together, Lesser and I realized that the questions asked of those solutions were quite different in biology from those in the physical chemistry he knew. Biology is inherited. Information is passed by handfuls of atoms in the genetic code. A few atoms in the proteins built from the code change macroscopic function. Indeed, a few atoms often control biological function in the same sense that a gas pedal controls the speed of a car. Biological questions then are most productive when they are asked in the context of evolution. What function does a system perform? How is the system built to perform that function? What forces are used to perform that function? How are the modules that perform functions connected to make the machinery of life. Physiologists have shown that much of life is a nested hierarchy of devices, one on top of another, linking atomic ions in concentrated solutions to current flow through proteins, current flow to voltage signals, voltage signals to changes in current flow, all connected to make a regenerative system that allows electrical action potentials to move meters, under the control of a few atoms. The hierarchy of devices allows macroscopic properties to emerge from atomic scale interactions. The structures of biology create these devices. The concentration and electrical fields of biology power these devices, more than anything else. The resulting organisms reproduce. Evolution selects the organisms that reproduce more and thereby selects the devices that allow macroscopic control to emerge from the atomic structures of genes and proteins and their motions.

Living systems pose a particular challenge for physical scientists. The goals of research are different when systems are living and that reality is not often apparent to physical scientists as they approach biological systems. Living systems are like engineering systems. They are built for a purpose, and they are interesting and useful to study only as they perform that purpose.

As I worked with Lesser Blum, it became clear that the reality of different goals was the essential issue separating physical and biological scientists.

Lesser had spent much of his life studying concentrated salt solutions well known to biologists as 'Ringer solutions'[1-6]. All life occurs in concentrated salt solutions, derived from seawater, whether the Ringer solutions outside cells or the concentrated salt solutions inside cells, or the even more concentrated salt solutions in and near the protein molecules that perform so many of life's functions [7-9]. Life is very much applied physical chemistry of concentrated salt solutions, but the questions we ask of living systems differ from those physical scientists ask, for the most part, as Lesser and I learned working together.

Life is about living systems as they are presented to us by evolution and questions must be asked in that context if questioning is to be efficient and answers are to be productive. Life is not about everything organic or biochemical molecules might do. Biological sciences are about what living systems actually do.

It is much easier to study what systems do when one has some idea of their design principle. What are the principles that govern the design of biological systems by evolution?

**Living systems are built to reproduce.** Living systems have a purpose and a function: to reproduce. The study of living systems needs always to remember that fact. This idea is the essence of evolutionary thinking "without which none of biology makes sense", paraphrasing the evolutionary biologist Dobzhansky [10]. Animals that reproduce more often soon come to dominate the population: systems that allow more reproduction replace less successful systems and life evolves so that more and more of the population has the (version of the) system that allows more reproduction. The living systems biologists study are those that have come to dominate populations of animals because they are more successful (in this reproductive sense). The systems biologists study are those that allowed more reproductive success at some crucial time in evolutionary history.

Studying living systems is similar to studying engineering systems in many ways because both consist of devices designed for a purpose. Both need to be studied as they (try to) fulfill that purpose. But studying living systems is different from engineering as well. The designers are very different in the two sciences. Engineers have logical purposes—if one includes making a profit as logical. Evolution has a general purpose, as we have seen, to facilitate reproduction, but we do not know how that plays out, or played out in the past when the system under consideration actually evolved. We do not know how the biological system contributed to reproductive success ***when it evolved*** and so we do not know the specific evolutionary purpose of a biological system. We have no access to the plans of evolution, or to enough historical information to know why evolution selected the adaptations that it did to make animals reproduce more effectively.

In fact, we are often confronted with a living system without even knowing what it does. (If this statement produces skepticism, look up the function of the cerebellum.) Then scientists must simply investigate and do everything we can think of, until we stumble on the function, and (if we are lucky) on its evolutionary role, and we can focus our work.

No one should dismiss this trial and error, exhaustive investigation just because it is exhausting. In fact, many medical advances of immediate importance to us and our families have been discovered this way as reference to the immense literature of 'drug design' will confirm. Most of pharmacological science and drug design is the systematic development of trial and error science, judging by costs and allocation of resources. In the presence of great total ignorance, one can make progress with reasonable efficiency when the needs are great enough, as they are in the design of drugs of improve our daily life. The essential idea is to design efficient assays to evaluate results, to determine if a change motivated by experiments or clinical trials is an error or success. If such assays can be designed, trial and error drug development can proceed effectively in many cases where we know very little of mechanism. We start with profound ignorance in many cases and we must do what we can because of the importance of biological research to everyday life and health of our families.

Trial and error is something we do when we feel we must do something.

**Studying biology is a kind of reverse engineering**, in which we must determine what the living system does, what its function is, and then proceed to study how the device performs the function. Studying how a living system performs a function is a kind of reverse engineering, determining the properties of a system from the outputs of the system. Reverse engineering is an inverse problem, and like other inverse problems, the results are often ill posed, [11, 12] very sensitive to the type and accuracy of data available. Asking the right question is crucial in studying inverse problems.[13, 14] Asking the right question is crucial in studying biology.

Reverse engineering can help us understand what questions to ask, and just as importantly what questions not to ask. Consider a simple amplifier. When it is functioning as it is designed, the amplifier multiplies an input voltage by a constant, and produces an output that depends only on the input. Of course, an amplifier can amplify in this way only in a certain range of conditions, when it is turned on, for example. Turning on the amplifier means connecting it to an appropriate power supply. If the power supply is turned off, the amplifier no longer multiplies by a constant. But the amplifier does not explode or disappear just because it is turned off. A turned off amplifier is an amplifier without a power supply. The turned off amplifier has a definite relation of input and output, but that relation is not useful because it does not follow a simple 'law'. The input output relation of a turned off amplifier is very complicated, and of little interest except to engineers studying device failure, even though it exists. Sadly, much of biochemistry is done with power supplies turned off, namely at equilibrium, without the nonequilibrium flows that energize life, as important as the nonequilibrium flows that energize amplifiers. Life at equilibrium has a name. Death.

It is clear we want to study an amplifier as it is used, when it is turned on. We must learn where to connect power, and what voltages supply that power, before we can study the amplifier intelligently. Once we know the location of the power connections, and the voltages they need, the inverse problem of reverse engineering is very easy, for an amplifier, that is. We can measure the input and output once, and determine their ratio, which is the amplifier's gain.

But locating the power connections, and determining the proper voltages for the power supply, is itself an inverse problem. This information is crucial. Without knowledge of the power needs of the amplifier, we can do almost nothing useful. With that information we can do a great deal.

Of course, one way to find the power needs of the amplifier is simply to try everything, in trial and error experimentation, taking every wire that comes out of the amplifier and seeing which are inputs, outputs, and power supplies by some experimental criteria or other. That can eventually work, in some cases, but probably not in every case.

The difficulties in reverse engineering are brought to our attention when we realize how easy it is to damage the system while testing it to see if wires are inputs or power supplies. Testing can easily damage the system. Applying a power supply voltage of say 5 volts to some input connections of the amplifier can have dramatic irreversible results in microseconds. Inputs are often designed for voltages of millivolts. Exposure to 5 volts can destroy them. The damage occurs very quickly and measurements take a long time, if one includes the time to set up the measurement. Finding the properties of an amplifier by trial and error measurement of a system so quickly destroyed by a measurement is a daunting task.

The lessons for biology are clear and are known to every biologist working in a laboratory. Conditions are crucial and must be maintained close to those in life. In some sense, every biological experiment is performed on a “preparation” of a living system.

Learning how to prepare the biological system and turn it into a useful, reproducible preparation, that shares the important properties of systems in animals, is one of the major tasks of biologists and is not often explained to physical scientists. Biologists proceed by isolating parts of systems that have reproducible properties and try to reconstruct the behavior of the entire system from those parts. The analogy with engineering is clear. Engineering systems need to be studied as they are used. So do biological systems. Otherwise they are much harder to understand, or even impossible. And who wants to understand systems that do not function anyway? Few want to study dead systems, although gross anatomy does have its uses.

**Molecular Biology** has had remarkable success because molecules are special parts of biological systems. Not all parts of a biological system have similar importance because not all parts of life have a similarly important role in reproduction.

Evolution, and thus life, is about reproduction. In fact (nearly) all the information needed for reproduction is carried in a handful or two of atoms called genes. Genes are coded into the sequence of nucleic acids by codons, (essentially a sequence of three base pairs of DNA). The genome is a special part of biology much more important than other parts because everything else in biology can be built from the plan coded in the DNA.

The only thing DNA can make is proteins. Proteins then in turn make everything else in biology, including other types of biological molecules (e.g., lipids, weak acids and bases, and so on), including the structures of life, most of which are made of protein (along with lipids for example).

The analogy with engineering is again useful and surprisingly accurate. DNA is the blueprint of life, as blueprints are the plans of buildings. Blueprints are much easier to change than buildings. DNA is much easier to change than proteins. Much more than blueprints are needed to construct a building, but the crucial information is in the plans. Much more than DNA is needed to construct a cell, but the crucial information is in the DNA and the genes it encodes. For this reason, the technology of DNA manipulation has received great attention and has become an engineering industry of economic importance.[15]

**DNA has all of the information of biology in sequences of nucleobases.** That is the first fact of molecular biology. **These genes make proteins.** That is the second fact of molecular biology. These proteins then make all of biology, by making other biomolecules (like fats and sugars), and molecular structures (like the structures of cells and tissues, from membranes to muscles to the whole heart).

In a very real sense, the proteins of biology are the main components of the parts list of life. Before a physical scientist like Lesser Blum tries to understand a biological system, he/she must have a list of parts and as much information about the parts as possible. That is why nearly all traditional biology is descriptive. As a foundation for understanding function, the identification, naming and understanding of parts is essential. The classical biological disciplines of anatomy, histology, biochemistry, and much of physiology are devoted to compiling a parts list. Biology is again like engineering. Before working on an amplifier, let alone a car, one must have a list of parts and as much information about the parts as possible. Engineering is inconceivable without a parts list. The first task of reverse engineering must be to compile a parts list so we know what we are dealing with.

Molecular and cell biologists have identified and analyzed the parts list of life with great success.[16] The proteins of life are known in great detail. In an astonishing triumph of science, structural biologists now know the position of individual atoms of more than 100,000 proteins. (Indeed, biologists know the positions of atoms and charges carried by some of those atoms with greater precision than physical scientists know the positions of atoms and their charges in the transistors of our computers.[17])

Molecular biologists are not content with just knowing the parts. They are fine experimental scientists who like to tinker with the parts they find. Many tools are available to change the blueprints of life, to change the DNA that codes for proteins. By modifying the blueprint, molecular biologists can change the proteins that are the parts list of life. A large fraction of modern molecular biology has learned to modify the blueprint of life—the genome—so that a handful of atoms in proteins can be changed at will, using commercially available kits of reagents. Changes in a handful of atoms are made routinely in thousands of laboratories every day using a technique called ‘site directed mutagenesis’ and its cousins.

**Changes in a handful of atoms of a protein often control biological function.** That is the third fact of molecular biology. Changes on the atomic scale often control macroscopic function in as definite a way as a gas pedal controls the speed of a car. The trial and error approach often works. Simply checking all possible modifications of DNA (really of the amino acids that make a protein) often produces most useful information about how the protein works, although of course this approach is always very hard tedious work and often does not succeed.

A central question of biology for a physical scientist then is to take the remarkable results of molecular biology and find out why they work. After the molecular biologist has shown that a handful of atoms controls biological function the physical scientist then asks **How do the atoms control biological function?**

A culture gap looms as we address this question. For biologists, it seems obvious that changing an acid group in a protein that carries a 'permanent' negative charge (e.g., the carboxylates on a glutamate side chain of the polypeptide chain that makes a protein) to a neutral group (changing a charged carboxylate in glutamate to an uncharged amine in glutamine) should have a large effect. Every biologist knows that biological function is controlled by the chemical properties of the amino acids that make up the polypeptide chain of proteins and changing an acid to a nonacid is a large change in chemical properties. Every biologists knows that the chemical properties of acidic and basic amino acids are determined by three atoms of water, and two atoms of the carboxylate.

**It is hard for a handful of atoms to control biological function.** What biologists do not always realize is that it is not easy or natural for a handful of atoms to control macroscopic function from a physical perspective. In fact, it is a miracle that a handful of atoms can control anything on a macroscopic scale if one considers the enormous number of atoms in a macroscopic system that do not control anything at all. (Think of the atoms in a door. Most of the atoms control nothing. But the atoms in the lubrication of the hinge and lock and key have an enormous effect and can control macroscopic function. These lubricant atoms exert control because they are in the right place. The structure of the door, or rather of the hinge and lock are built so the lubricant is important.) **A handful of atoms can control macroscopic function only because the structures of biology are built to make that happen**, as we shall now discuss in detail.

Atoms are angstroms in diameter, $10^{-10}$ meters, 0.1 nm. The smallest biological structures are proteins typically made of hundreds of thousands of atoms. Many biological structures are on the cellular scale of $10^{-5}$ meters, and most biological structures are much larger than that. They are on the millimeter to meter scale. The number of atoms involved in biological structures is very much larger (say $10^{17} \times$ larger) than the number of atoms in the chemical structures that control those structures.

Statistical mechanics[18-21] is the science that links the properties of atoms with macroscopic systems. Statistical mechanics is quite successful at averaging the properties of ideal gases and other systems with minimal interactions. Yet it is also obvious that the ordinary kind of averaging in statistical mechanics is insensitive to the properties of a tiny fraction of atoms. We [22] (Section 3.1, p. 22) actually performed this kind of average explicitly in some detail to show how important is the assumption of minimal interactions in classical statistical mechanics. The controlling atoms in biological structures often make up less than $10^{-17}$ part of the structure, so the physical scientist finds it hard to believe that a handful of atoms can control macroscopic function! The physical scientist knows that a continental country (think Russia, USA, Canada, Australia) has some $10^{13}$ sq. meters in it. No conceivable change to 1 square meter is likely to change the average properties of Russia, Canada, or the USA. Even the explosion of a nuclear weapon (of 1 sq meter cross section) is unlikely to change the average properties of a continent very much. But the atoms controlling biology are a much smaller part of an organism than 1 square meter is of a continent. No wonder the physicist is skeptical of atomic control of biological function.

**The biologist considers obvious what the physical scientist has difficulty believing**.

The averaging procedure of the physicist does not give insight into how a few atoms controls macroscopic function. The averaging procedure of the physicist does not deal with strongly correlated signals and systems. Yet the input and output of an amplifier, or a device for that matter, are perfectly correlated, with coherence function of unity![23] I believe biology uses a hierarchy of devices to connect atomic structure and macroscopic function. Each of those devices has a coherence function close to one. Each of those devices has highly correlated inputs and outputs. Those correlations are what allow the macroscopic properties of an organism to emerge from atomic detail. Those correlations are what are left out in traditional averaging methods of statistical physics.

**Biology depends on structure, on all scales.** The answer to this apparent paradox lies in the structure of the biological system (Fig. 1). The biological system is constructed with

Fig. 1

**Hierarchy of Life**

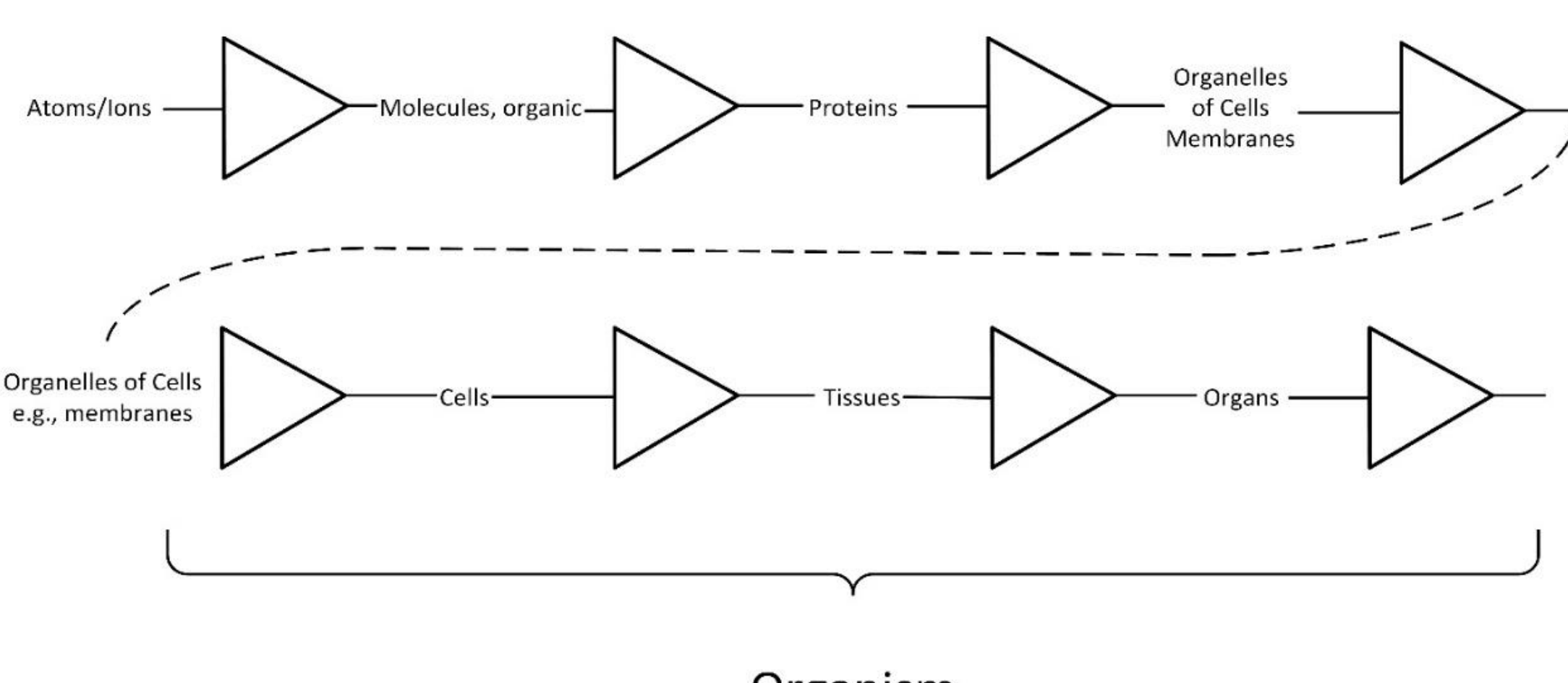

a very special structure that allows handfuls of atoms to control macroscopic function. The properties of this structure do not violate physical laws at all. Rather the structure has **'emergent properties'**, as they are called to the confusion of my mathematician friends. Use of the adjective 'emergent' confuses the idea, particularly when non-native speakers are struggling with 'emerging', 'emergent', instead of just thinking of the simple verb 'emerge'. The idea is simple. Some properties emerge (i.e., are visible or even exist) only on certain scales. Amplifiers cannot be seen to amplify if they are observed only with atomic resolution. Animals cannot be seen to reproduce if they are observed only on a cellular or molecular or atomic scale.

**Emergent Properties** are properties that emerge on some scales, but are hard to see on others. The properties emerge in biology because the systems were built so they would emerge. Evolution builds systems so their macroscopic properties are controlled by handfuls of atoms called genes. It can do nothing else. Everything in biology is inherited and the only thing that is inherited are genes. Without emergent properties, life could not exist because it could not be inherited. The macroscopic properties of life emerge from the astronomical number of atoms and their larger than astronomical number of interactions because the structures of biology are built (by evolution) to make them emerge!

There is nothing mysteriously biological about such emergent properties. We all know that a door handle opens a door because of its structure, and its connections, i.e. its correlated motions. The properties of the metal it is made from have little to do with its function. The atomic scale properties of the metal are even less important. Evolution does nothing more than engineers. Engineers sculpt metal into a specific structure that is the handle of a door. Biology sculpts atoms and molecules into cells and tissues and organisms with a specific function.

Biological systems are like engineering systems. They are built with structures that give them special properties. Averaging atoms is a poor way to find these structures or detect their roles, although certainly averaging—like trial and error science—might be possible. Rather, the general scientific method allows one to detect these structures and their roles. One observes structure and function at different scales, and builds models of those structures and functions, including just what is needed to see how they work, and how they interact.

Using this process, physiologists, more than anyone else, have shown that biological function is often a nested hierarchy of devices. Physiologists have built models at each scale from the atoms of genes, to the atoms and moieties of proteins, to the proteins of cell membranes and cells themselves, to the cells of tissues, and the tissues or organisms. Each model is a module, usually a device, with a definite input and definite output. The outputs are usually at a larger scale than the inputs. The (nested) hierarchy of devices allows atoms to control macroscopic function.

The nested hierarchy is understood in nearly complete detail in the case of nerve signals (see Fig. 2).

**A hierarchy of devices allows atoms to control animals.** The crucial fact is that handfuls of atoms do control macroscopic biological function. Evolution has designed animals so a few atoms of their proteins control the biological functions crucial for the reproduction of the species. Indeed, the work of physiologists for more than a century [1-5] suggests that many, if not most of biology uses a simple plan to exert control from atoms to cells to tissues to organs to animals. Physiologists have shown that many systems in biology are modular and function as devices with definite outputs controlled by specific inputs according to robust rules. The outputs are typically on a large scale than the inputs so a few devices in series serves to link the smallest input (say the atoms of the voltage sensor of channel [24] or of a selectivity filter [25, 26] to macroscopic function. The atoms of the selectivity filter control the current through the open channel in one device. The atoms of the voltage sensor control how often the channel is open in another device. Together they form the device of a channel which controls the current through the membrane of a cell. That current in turn controls the spread of electrical potential on the millimeter scale through a device which is in fact a cable equation, the familiar telegrapher's equation of partial differential equation theory. The spread of potential then controls the spatial activation (i.e., opening) of channels according to the Hodgkin Huxley of differential equations. This hierarchy of devices makes the action potential signal of nerves that spreads meters. Physiologists have thus developed a general approach to biological investigation.

**A productive working hypothesis for investigating biological systems is seen here. It is useful to assume that evolution has built a nested hierarchy of devices that allow a few atoms to control biological function.[27-31]** The scientific process then creates models of these devices and their connections. Mathematics predicts the behavior of the hierarchy of models. Experiments check these models.

Fig. 2

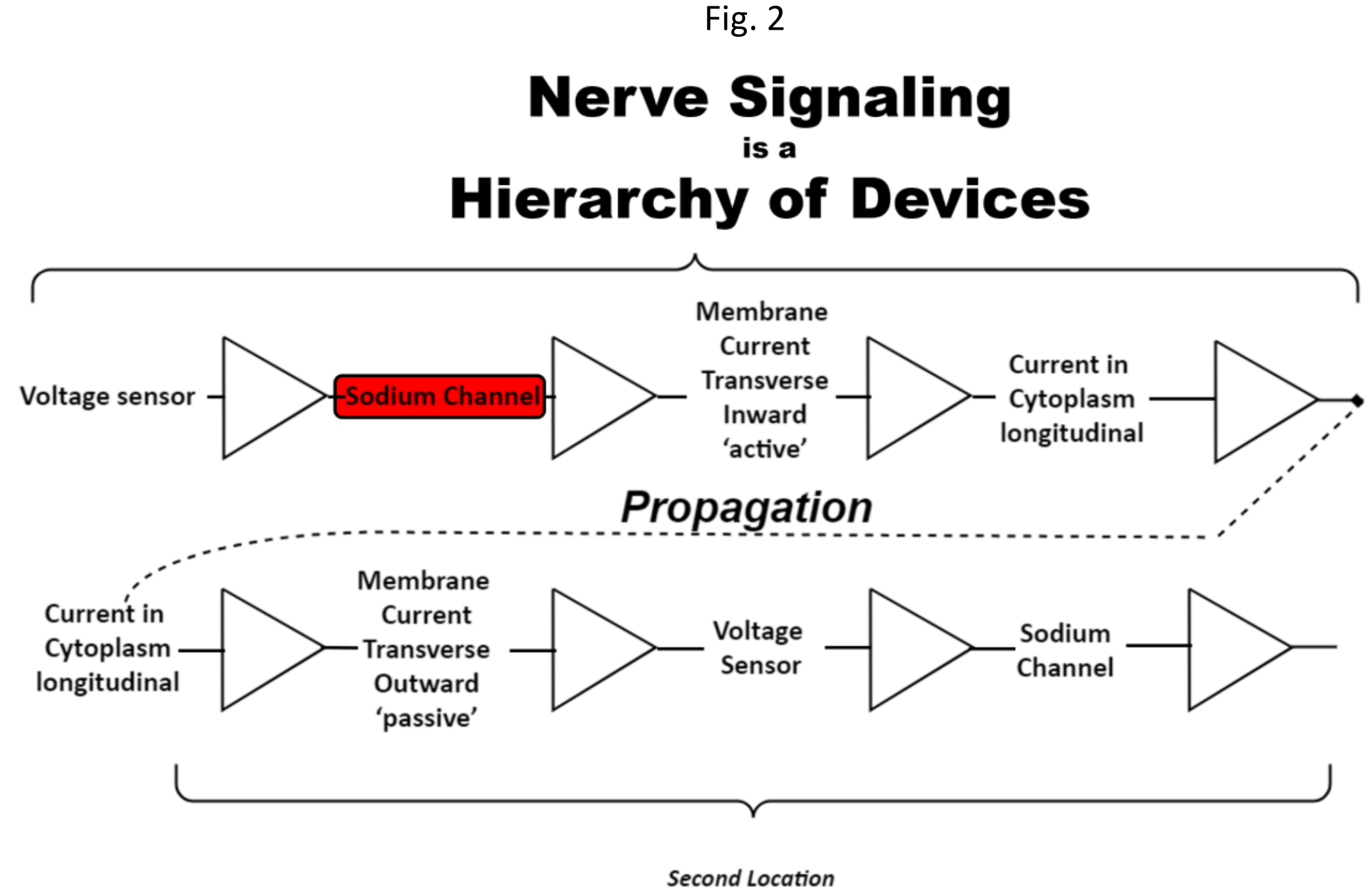

detail (diameter). Life depends on atomic detail.

The migration and thermal motion of these atomic ions produces macroscopic current flow that links the properties of atoms, controlled by the channels (that are protein macro-molecules), to the nerve signal, linking the atomic scale of control, to the macroscopic signal that is the function of nerve axons, that is the action potential. The motions of atoms and macromolecular channels control the flow of current governed by the equations of electrodiffusion. The current flows according to a reduced version of Maxwell's equations, called the cable equation in biophysics,[32-36] and the telegrapher's equation in mathematics. The equations of electrodiffusion and current flow link the atomic structures and ions to the action potentials that propagate meters, from toe to spinal cord in mammals. Electrodiffusion and current flow link the atomic signals inherited from genes (in the form of nucleic acids) to the action potentials that control the life of animals on the macroscopic scale.

Determining the hierarchy of devices involved in the nerve signal took extraordinary efforts of hundreds of laboratories and thousands of scientists recognized in many Nobel Prizes, that themselves form a hierarchy of knowledge, from Lord Adrian, to Hodgkin and Huxley, to Bernard Katz, Neher and Sakmann, and MacKinnon, among others. At each level of the hierarchy, biologists had to learn what was crucial and what was much less important. Adrian learned that the frequency of signals was crucial, not their size. Hodgkin learned that the flow of current was crucial to the propagation of action potentials [37, 38], not chemical diffusion (as was thought by his elder contemporaries [39]. Cole, Hodgkin and Huxley learned that voltage controlled the channels of the axon (more than anything else). Katz learned how a chemical controlled a channel crucial for cell to cell transmission of information. and so on.

What is crucial to understand is that at each stage of this hierarchy of knowledge, the choice of questions was at least as difficult as answering the question. The great biologists learned how to choose the right questions. The answers to these questions allowed one to understand how each process fit into the hierarchy of life as well as the hierarchy of knowledge. Physical scientists and mathematicians are much less likely to know the right questions than biologists.

**A crucial part of physical analysis of biological systems is the choice of questions: the questions are developed by experimental biologists who are experts on that system and its biological role.** Mathematicians study the questions those biologists think are important.

The structures form the skeleton of our models (in mathematical language they provide crucial constraints). We believe it is the role of the mathematician to add flesh (actually muscle) to that skeleton so it can move.

The role of structure in biology is a main theme well known to the hundreds of thousands of scientists doing molecular biology every day. What is not so well known is the necessary role of mathematics, in my view, and so I enumerate it here in detail I fear will be painful to read by the physical scientists for whom the role of mathematics is obvious.

I believe mathematics is needed to

1) Describe the functions of the modules and devices of life.

2) Show how the structure imposed by life, from atoms to cells, to tissues, to organs, constrains and creates function.

3) Show how physical laws are constrained, even channeled by structure, and so produce function. In other words, to show how inheritable information (i.e., genes and individual amino acids) control function, by executing physical laws in defined structures.

5) Test the hypothesis that evolution creates devices, and links them in a nested hierarchy so handfuls of atoms can control macroscopic function, although biological motions are some $10^{13}\times$ slower than atomic motions, and atomic sizes are $10^{7}\times$ smaller than biological cells.

6) Create computable models that can be compared with experimental data from biological systems already studied in great detail, because of their crucial role in the hierarchy of life.

7) Design experiments to compare models. Competing models are hallmarks of important science. Distinguishing between models is a crucial step in scientific progress. Theory and simulation can design experiments to distinguish between theories and models. These can be explored ahead of time in models and so save enormous amounts of experimental time, because theory and simulations can use computation instead of trial and error experimentation, in favorable cases.

8) Design new experiments to reveal natural function. Theory and simulations may reveal ways to explore natural function that have not yet been thought of. The measurement of gating current is one example that comes to mind. Without a theory in mind [40, 41], it is hard to imagine experimentalists seeking a nonlinear displacement current that is <1% of ionic current!

This is the plan my collaborators and I have tried to follow in understanding the role of ion channels in biological function.[9, 42-51] The ions of ion channels are concentrated in the solutions outside of cells and enormously further concentrated inside the channels themselves, because of the large charge density on the walls of the channel's pore.

These concentrated ions need to be understood in the language of physics and chemistry that Lesser Blum taught me. The biological role of the concentrated ions needs to be understood in the context of the function of channels, and their structure, as has been studied by innumerable biologists, with remarkable success in the last century or so.

The biologists ask the questions, provide the parts list, and the structures. The physical scientists are needed to show how these structures move. The mathematicians are needed to show how the movement of these structures answers the questions of biologists, and of life.

## References

[1] W. Boron, E. Boulpaep, Medical Physiology, Saunders, New York, 2008.
[2] T.C. Ruch, H.D. Patton, Physiology and Biophysics, Volume 3: Digestion, Metabolism, Endocrine Function and Reproduction, W.B. Saunders Company, Philadelphia, 1973.
[3] T.C. Ruch, H.D. Patton, Physiology and Biophysics, Volume 1: The Brain and Neural Function, W.B. Saunders Company, Philadelphia, 1973.
[4] T.C. Ruch, H.D. Patton, Physiology and Biophysics, Volume 2: Circulation, Respiration and Balance, W.B. Saunders Company, Philadelphia, 1973.
[5] C.L. Prosser, B.A. Curtis, E. Meisami, A History of Nerve, Muscle and Synapse Physiology, Stipes Public License2009.
[6] N. Sperelakis, Cell physiology source book: essentials of membrane biophysics, Elsevier2012.
[7] D. Jimenez-Morales, J. Liang, B. Eisenberg, Ionizable side chains at catalytic active sites of enzymes, European Biophysics Journal, 41 (2012) 449-460.
[8] B. Eisenberg, Proteins, Channels, and Crowded Ions, Biophysical chemistry, 100 (2003) 507 - 517.
[9] B. Eisenberg, Crowded Charges in Ion Channels, in: S.A. Rice (Ed.) Advances in Chemical Physics, John Wiley & Sons, Inc., New York, 2011, pp. 77-223 also on the arXiv at http://arxiv.org/abs/1009.1786v1001.
[10] T. Dobzhansky, Nothing in Biology Makes Sense Except in the Light of Evolution, The American Biology Teacher, 35 (2013) 125-129.
[11] H.W. Engl, M. Hanke, A. Neubauer, Regularization of Inverse Problems Kluwer, Dordrecht, The Netherlands, 2000.
[12] J. Kaipio, E. Somersalo, Statistical and Computational Inverse Problems Springer, New York, 2005.
[13] M. Burger, Inverse problems in ion channel modelling, Inverse Problems, 27 (2011) 083001.
[14] M. Burger, R.S. Eisenberg, H. Engl, Inverse Problems Related to Ion Channel Selectivity, SIAM J Applied Math, 67 (2007) 960-989
[15] J.D. Watson, A. Berry, K. Davies, DNA: The Story of the Genetic Revolution, Knopf2017.
[16] B. Alberts, D. Bray, J. Lewis, M. Raff, K. Roberts, J.D. Watson, Molecular Biology of the Cell, Third ed., Garland, New York, 1994.
[17] D. Vasileska, S.M. Goodnick, G. Klimeck, Computational Electronics: Semiclassical and Quantum Device Modeling and Simulation, CRC Press, New York, 2010.
[18] S.R. Berry, S.A. Rice, J. Ross, Physical Chemistry, Second Edition ed., Oxford, New York, 2000.
[19] S.A. Rice, P. Gray, The Statistical Mechanics of Simple Fluids, Interscience (Wiley), New York, 1965.
[20] J.-P. Hansen, I.R. McDonald, Theory of Simple Liquids, Third Edition ed., Academic Press, New York, 2006.
[21] J.-L. Barratt, J.-P. Hansen, Basic concepts for simple and complex liquids, Cambridge University Press2003.
[22] B. Eisenberg, X. Oriols, D. Ferry, Dynamics of Current, Charge, and Mass, Molecular Based Mathematical Biology 5(2017) 78-115 and arXiv preprint https://arxiv.org/abs/1708.07400.

[23] J.S. Bendat, A.G. Piersol, Random data: analysis and measurement procedures, John Wiley & Sons2011.
[24] T.-L. Horng, R.S. Eisenberg, C. Liu, F. Bezanilla, Gating Current Models Computed with Consistent Interactions, arXiv preprint arXiv 1707.02566, (2017)
[25] D. Boda, W. Nonner, D. Henderson, B. Eisenberg, D. Gillespie, Volume Exclusion in Calcium Selective Channels, Biophys. J., 94 (2008) 3486-3496.
[26] D. Boda, M. Valisko, D. Henderson, B. Eisenberg, D. Gillespie, W. Nonner, Ionic selectivity in L-type calcium channels by electrostatics and hard-core repulsion, Journal of General Physiology, 133 (2009) 497-509.
[27] B. Eisenberg, Ion Channels as Devices, Journal of Computational Electronics, 2 (2003) 245-249.
[28] B. Eisenberg, A Leading Role for Mathematics in the Study of Ionic Solutions, SIAM News, 45 (2012) 11-12.
[29] B. Eisenberg, Life's Solutions. A Mathematical Challenge. , Available on arXiv as http://arxiv.org/abs/1207.4737, DOI (2012).
[30] B. Eisenberg, Living Devices: The Physiological Point of View, Available on arXiv as http://arxiv.org/abs/1206.6490, DOI (2012).
[31] B. Eisenberg, Living Transistors: a Physicist's View of Ion Channels, available on http://arxiv.org/ as q-bio/0506016v2 24 pages, DOI (2005).
[32] L. Kelvin, On the theory of the electric telegraph, Philosophical Magazine, 11 (1856) 146-160.
[33] L. Kelvin, On the theory of the electric telegraph, Proceedings of the Royal Society (London), 7 (1855) 382-399.
[34] L.D. Davis, Jr.,, R.L. de No, Contribution to the Mathematical Theory of the electrotonus, Studies from the Rockefeller Institute for Medical Research, 131 (1947) 442-496.
[35] J.J.B. Jack, D. Noble, R.W. Tsien, Electric Current Flow in Excitable Cells, Oxford, Clarendon Press., New York, 1975.
[36] V. Barcilon, J. Cole, R.S. Eisenberg, A singular perturbation analysis of induced electric fields in nerve cells, SIAM J. Appl. Math., 21 (1971) 339-354.
[37] A.L. Hodgkin, Evidence for electrical transmission in nerve: Part II, J Physiol, 90 (1937) 211-232.
[38] A.L. Hodgkin, Evidence for electrical transmission in nerve: Part I, J Physiol, 90 (1937) 183-210.
[39] A.V. Hill, Chemical Wave Transmission in Nerve, Cambridge University Press1932.
[40] A. Hodgkin, A. Huxley, B. Katz, Ionic Currents underlying activity in the giant axon of the squid, Arch. Sci. physiol., 3 (1949) 129-150.
[41] A.F. Huxley, From overshoot to voltage clamp, Trends in neurosciences, 25 (2002) 553-558.
[42] B. Eisenberg, Can we make biochemistry an exact science?, Available on arXiv as https://arxiv.org/abs/1409.0243, DOI (2014).
[43] B. Eisenberg, Shouldn't we make biochemistry an exact science?, ASBMB Today, 13 (2014) 36-38.
[44] B. Eisenberg, Ionic Interactions Are Everywhere, Physiology, 28 (2013) 28-38.

[45] B. Eisenberg, Ions in Fluctuating Channels: Transistors Alive, Fluctuation and Noise Letters, 11 (2012) 1240001 available on arXiv.org with Paper ID arXiv:q-bio/0506016v1240003.
[46] B. Eisenberg, Multiple Scales in the Simulation of Ion Channels and Proteins, The Journal of Physical Chemistry C, 114 (2010) 20719-20733.
[47] B. Eisenberg, Engineering channels: Atomic biology, Proceedings of the National Academy of Sciences, 105 (2008) 6211-6212.
[48] R.S. Eisenberg, From Structure to Function in Open Ionic Channels, Journal of Membrane Biology, 171 (1999) 1-24, available on arXiv at https://arxiv.org/abs/1011.2939.
[49] R.S. Eisenberg, Atomic Biology, Electrostatics and Ionic Channels., in: R. Elber (Ed.) New Developments and Theoretical Studies of Proteins, World Scientific, Philadelphia, 1996, pp. 269-357. Published in the Physics ArXiv as arXiv:0807.0715.
[50] R.S. Eisenberg, Computing the field in proteins and channels., Journal of Membrane Biology, 150 (1996) 1–25. Also available on http:\\arxiv.org as arXiv 1009.2857.
[51] R.S. Eisenberg, Channels as enzymes: Oxymoron and Tautology, Journal of Membrane Biology, 115 (1990) 1–12. Available on arXiv as http://arxiv.org/abs/1112.2363.